\documentclass[useAMS,usenatbib,letterpaper]{mn2e}
\usepackage{amssymb,amsmath,epsfig,times,natbib}

\usepackage{aecompl}
\usepackage{lmodern}
\long\def\symbolfootnote[#1]#2{\begingroup%
\def\thefootnote{\fnsymbol{footnote}}\footnote[#1]{#2}\endgroup} 

\title[PCA of MCG--06-30-15]{Principal Component Analysis of MCG--06-30-15 with \emph{XMM-Newton}}

\author[M. L. Parker et al.]{M. L. Parker,$^1$\thanks{Email: 
    mlparker@ast.cam.ac.uk}  A. Marinucci,$^{2,3}$ L. Brenneman$^4$, A. C. Fabian,$^1$  E. Kara,$^1$
    G. Matt,$^2$   \newauthor and D. J. Walton$^5$\\
  $^1$Institute of Astronomy, Madingley Road, Cambridge CB3 0HA \\
  $^2$Dipartimento di Matematica e Fisica, Universit\`{a} degli Studi Roma Tre, via della Vasca Navale 84, 00146 Roma, Italy \\
  $^3$Centro de Astrobiolog\'ia (CSIC-INTA), Dep. de Astrof\'isica; LAEFF, Villanueva de la Ca$\tilde{n}$ada, Madrid, Spain\\
  $^4$Harvard-Smithsonian Center for Astrophysics, 60 Garden St, Cambridge, MA 02138, USA \\
  $^5$California Institute of Technology, 1200 East California Boulevard, Pasadena, CA 91125, USA \\
}
\date{}

\begin{document}

\maketitle

\begin{abstract}
We analyse the spectral variability of MCG--06-30-15 with 600~ks of \emph{XMM-Newton} data, including 300~ks of new data from the joint \emph{XMM-Newton} and \emph{NuSTAR} 2013 observational campaign. We use principal component analysis to find high resolution, model independent spectra of the different variable components of the spectrum. We find that over 99 per cent of the variability can be described by just three components, which are consistent with variations in the normalisation of the powerlaw continuum ($\sim$ 97 per cent), the photon index ($\sim2$ per cent), and the normalisation of a relativistically blurred reflection spectrum ($\sim0.5$ per cent). We also find a fourth significant component but this is heavily diluted by noise, and we can attribute all the remaining spectral variability to noise.\\
All three components are found to be variable on timescales from 20~ks down to 1~ks, which 
corresponds to a distance from the central black hole of less than 70 gravitational radii. We compare these results with those derived from spectral fitting, and find them to be in very good agreement with our interpretation of the principal components.\\
We conclude that the observed relatively weak variability in the reflected component of the spectrum of MCG--06-30-15 is due to the effects of light-bending close to the event horizon of the black hole, and demonstrate that principal component analysis is an effective tool for analysing spectral variability in this regime.
\end{abstract}


\section{Introduction}
MCG--06-30-15 is a Seyfert 1 galaxy, with a highly variable central X-ray source.
It has been very well studied, and shows a strong, broad iron line feature \citep{Tanaka95}, along with an
excess at soft energies, and a Compton hump at high energies \citep{Miniutti07}. This is indicative of a relativistic 
reflection spectrum, caused by the primary coronal emission hitting the accretion disk and causing fluorescent line emission.
This spectrum is then blurred by relativistic effects, close to the event horizon.

\begin{figure}
\centering
\includegraphics[width=\linewidth]{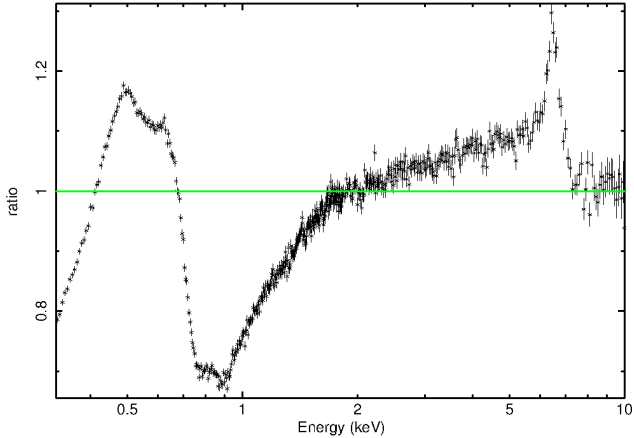}
\caption{Ratio of the 2013 MCG--06-30-15 XMM spectrum to a powerlaw, fit in the 1.9--2 and 9--10 keV bands.
The response is modified between 2 and 2.5 keV using `gain fit' in \textsc{Xspec}, to correct for the instrumental 
gold edge (purely for visual purposes). A strong narrow iron line is visible 
between 6 and 7 keV, along with a broad iron line stretching to lower energies, a soft excess around 0.5 keV, and complex absorption features below 2 keV. Some 
binning is applied, for clarity.}
\label{xmmratio}
\end{figure}

MCG--06-30-15 also shows evidence of
complex absorption features, requiring multiple absorbing zones to describe them fully 
\citep{Otani96,Lee01,Turner03,Young05,Chiang11}. Fig.~\ref{xmmratio} shows the ratio of
the \emph{XMM-Newton} spectrum from the latest observation of MCG--06-30-15 to a powerlaw, fit 
in the 1.9--2 and 9--10keV energy bands. This clearly shows the effect of absorption,
as well as the soft excess and broad and narrow iron lines.

The nature of the variability and spectral features in MCG--06-30-15 is a contentious issue. The red wing
of the iron line, instead of being symptomatic of relativistic blurring of a reflection spectrum, was instead interpreted 
as a sign of partial covering absorbers by \cite{Miller08}. In this model, a fraction of the central X-ray source
is covered by an additional absorber. In combination with a distant 
reflector, which is also present in the blurred reflection model, this explains the broad line feature.

A key feature of the emission from MCG--06-30-15 is the lack of variability in the non-powerlaw component \citep{Fabian02}.
This component of the spectrum appears to stay relatively constant, while almost all of the variability can be explained
by a change in the normalisation of the powerlaw. Initially, this is hard to understand in the context of either model - if the 
constant component is due to reflection, an increase in the continuum flux should lead to an increase in the reflected flux and
if this component is due to partial covering of the powerlaw, the unabsorbed part of that powerlaw responsible
for the red wing and soft excess should track the powerlaw flux as well.

In the partial covering model, it appears that the covering fraction must be inversely proportional to the powerlaw flux \citep{Reynolds09},
which requires a change in the size of either the source or the absorber, fine-tuned so as to keep the flux in the absorbed component
constant. In the reflection scenario, a relatively constant reflection component can be explained by the light bending model of 
\citet{Miniutti03}, where the height of the source above the disk changes, producing a large change in the continuum flux
with very little change in the reflection component. In this model, the source height determines the continuum
flux, as the closer the source is to the black hole, the larger the fraction of source photons that pass through
the event horizon. The reflection component stays relatively constant because the fraction of source photons that hit 
the disk is not strongly dependent on the source height - the photons that would have hit the disk but are bent into 
the black hole are replaced by those that would have escaped to infinity, but are bent down onto the disk instead.

Time lags have been observed in MCG--06-30-15, by \citet{Emma11} and \citet{Vaughan03}. \citeauthor{Emma11} found a soft lag
between the 0.5--1.5 and 2--4~keV bands at frequencies around $10^{-3}$~Hz. They found that the lag-frequency spectrum was 
best described by a reverberation model in which the hard lag is due to variations propagating through the accretion disk and
the soft lag is caused by the light travel time between the corona and the disk, similar to that proposed for 1H0707-495
by \citet{Zoghbi11}.

To understand the emission of MCG--06-30-15, we need to consider variations from both spectral and timing perspectives, to constrain
both the variation timescales and the shape of the different components that make up the spectrum. In this paper, we use principal 
component analysis to simultaneously analyse the spectral shape and variability of the spectrum of MCG--06-30-15.

\section{Principal Component Analysis}

Principal Component Analysis (PCA) is a powerful tool for analysing variability
in complex sources, by factorising the dataset into a set of independently variable
components, expressing the maximum amount of variability in as few as possible.
This gives a model independent method of finding the different constituent parts of
a spectrum, and quantifying their variability.

PCA has applications in many areas of astronomy, and has been used for many different types of 
analysis, including quasar spectral parameters \citep{Francis99},
spectral variability of X-ray binaries \citep{Malzac06,Koljonen13}, UV spectral variability in AGN \citep{Mittaz90},
and stellar classifications \citep{Whitney83}. See \citet{Kendall75} for a discussion of the 
wider uses of this type of analysis outside astronomy.

\subsection{Theory}
PCA is effectively a method of reducing the dimensionality of a $n\times m$ 
dimensional dataset to a set of eigenvectors describing the majority of the variability
in the first few vectors. The method described here uses singular value decomposition (SVD), a method of decomposing a matrix
into orthogonal eigenvectors, to find the principal components \citep{Press86,Mittaz90,Miller07,Miller08},
as this has the advantage of not requiring a unique solution to the matrix factorisation. In practise, this means that it is
not necessary to have more spectra than energy bins in the analysis (analogous to having more unknowns than equations),
so the full instrumental resolution can be preserved, if desired.

We use SVD to reduce the number of values needed to describe a spectrum from $n$ parameters
(one for each energy bin, $E_\textrm{j}$) to a small number of principal components, which account for
the majority of the variability of the source. The remaining components are attributable
to noise, and can be distinguished from real components using the log-eigenvalue (LEV) diagram (see Fig.~\ref{LEV_10ks_1}).

We create a $n \times m$ matrix $M$, with $n$ energy bins and $m$ spectra from different
time bins. SVD is then used to factorise this matrix:
\begin{equation}
M=UAV^*
\label{matrixfactoring}
\end{equation}
where $U$ is a $n \times n$ matrix, $V$ is a $m \times m$ matrix, and $A$ is a $n \times m$
diagonal matrix.The rows of $U$ and the columns of $V$ each give a set of orthogonal eigenvectors to
the matrices $MM^*$ and $M^*M$, and the diagonal elements of $A$ are the corresponding eigenvalues.
When applied to a matrix of spectra, the eigenvectors give the spectra of each variable component, and
the square root of each eigenvalue quantifies the variability in that component. The fractional 
variability of each component can then be found by dividing each eigenvalue by the sum of all 
eigenvalues. By plotting the fractional variability of each eigenvalue on a log scale (see 
fig.~\ref{LEV_10ks_1}), the real variable components can be distinguished from the noise. In general,
the variability of the noise components decays geometrically, leading to a straight line on the 
LEV. Once identified, these eigenvectors can then be discarded from further analysis.

\subsection{Previous use of PCA with MCG6}
PCA of MCG--06-30-15 was first performed by \citet{Vaughan04}, using data from a long \emph{XMM-Newton}
observation in 2001. They found that 96 percent of the variability could be attributed to a single, relatively
flat, spectral component. This is consistent with their interpretation of the spectrum as a constant reflection
spectrum and highly variable powerlaw, and agrees with their results from RMS spectra and flux-flux
analysis. However, this analysis was limited to an energy range of 3--10~keV to exclude the potential effects of the warm absorber, and had low spectral
resolution, leaving the results ambiguous. The second and third components are shown, and are found 
to be consistent with noise.

More recently, \citet{Miller08} performed a more detailed analysis using SVD to preserve the full
instrumental resolution, with both \emph{XMM-Newton} and \emph{Suzaku} data. We note that it is not always desirable to use the full instrumental
resolution in studies of variability, as higher order terms may be lost from the analysis, and the noise in the component spectra is increased. \citeauthor{Miller08} find that the optimum signal to noise is achieved using 20~ks time bins, which are too large to examine the variability of the spectrum close to the event horizon.
In their analysis from 2-10keV they find a single variable component, well fit with a powerlaw. Below 2keV, they find that more components are necessary to fully describe the dataset, and attribute these to the effects of variable absorption, although they are not shown or modelled within that work.

\section{Data}

We use all the available \emph{EPIC-PN} \citep{Struder01} data for MCG6, including both the original 300ks used by 
\citeauthor*{Vaughan04} and \citeauthor*{Miller08}, which is publically available,
and $\sim$300ks from the recent joint \emph{XMM-Newton} \citep{Jansen01} and \emph{NuSTAR} \citep{Harrison13} 2013 observational campaign. 

We use a 40 arcsecond source extraction region,
and background regions of around 50 arcsec, and filtered the data for background flares.
See Marinucci et al. (in prep.) for a more detailed discussion of the
data reduction. The full list of observations used is shown in Table~\ref{obstab}.

\begin{table}
\centering
\begin{tabular}{l l l}
Observation ID & Date & Duration (s) \\
\hline
0111570101 & 2000-07-11 & 46453 \\
0111570201 & 2000-07-11 & 66197 \\
0029740101 & 2001-07-31  & 89432 \\
0029740701 & 2001-08-01 & 129367 \\
0029740801 & 2001-08-05 & 130487 \\
0693781201 & 2013-01-31 & 134214 \\
0693781301 & 2013-02-02 & 134214 \\
0693781401 & 2013-02-03 & 48918 \\
\hline
Total & & 779282
\end{tabular}
\caption{List of observations used in the PCA. Note that the on-source exposure times will be smaller than the on time, 
so the total exposure length is closer to 600ks.}
\label{obstab}
\end{table}

\subsection{Analysis}

We use custom good time interval (GTI) files to extract spectra with different time bins, splitting the data into $m=t_{total}/t_{bin}$ sections, and disregarding those with an on-source exposure time less than 30 per cent of the bin size. We then use \texttt{PyFits} 3.1.2 to read the spectra into \texttt{Python} for analysis.

For the PCA, we calculate a mean spectrum, $F_\textrm{mean}(E_\textrm{j})$, from all the background-subtracted individual spectra, and subtract this mean spectrum from each spectrum, giving a set of residual spectra.
These show the deviations from the mean for each spectrum, and these are then normalised by dividing by the mean number of counts in each bin, returning a set of fractional residual spectra:
\begin{equation}
F_\textrm{res,i}(E_\textrm{j})=\frac{F_\textrm{i}(E_\textrm{j}))-F_\textrm{mean}(E_\mathrm{j})}{F_\textrm{mean}(E_\mathrm{j})}
\end{equation}
An $n\times m$ matrix is then created from these, with $n$ energy bins and $m$ time bins. The energy bins are logarithmically spaced, and we vary the number of bins to optimise
the signal to noise ratio for each component, depending on the fractonal variance of the component currently being investigated
and the size of the time step used (lower variance means that the component has a smaller signal in the PCA, and smaller time bins
increase the noise). Finally, we use the \texttt{linalg.svd()} function from the 
\texttt{NumPy} library for \texttt{Python} on this matrix to calculate the SVD of the matrix.

We note that the PCA requires all components to be orthogonal. In practise, this means that the dot-product of any two spectra must be zero, i.e.:
\begin{equation}
\sum_{i=1}^n f_\textrm{a}(E_\mathrm{i})\times f_\textrm{b}(E_\mathrm{i})=0
\end{equation}
where $f_\textrm{a,b}$ are the component spectra. Given that the first component is greater than zero in all energy bins and approximately constant, this condition requires all subsequent principal components to have a mean value of approximately zero. This does not necessarily compromise the physical interpretation of the principal components, but means that only the most variable additive component should be expected to be entirely positive. Any other additive components will then be expressed as corrections to the first component.

We perform the PCA using several different timesteps: 20~ks, 10~ks, 5~ks, and 1~ks, and vary the number of spectral bins from 120 in the 20~ks analysis to 50 in the 1~ks analysis, over the energy range 0.4--9~keV. The results obtained are qualitatively the same, regardless of the size of the intervals used. The noise increases when shorter intervals or finer energy bins are used, and we find that 10~ks time intervals offer the best compromise in terms of signal to noise against timing resolution, so all figures presented use these intervals unless otherwise stated.

Increasing the energy range, and hence number of counts, greatly reduces the contribution of noise to the variance of the spectrum, but also exposes the analysis to more potential sources of variation, making the results more complex and harder to interpret. We restrict the analysis to the 0.4--9~keV band due to increased noise in the principal components at extreme energy values.

Random noise in the spectra can mostly be removed as higher-order components, and what remains can be estimated using various methods, such as perturbing the input spectra and examining the Log-Eigenvalue (LEV) diagram. Of more concern are systematic errors, which are harder to quantify. Because the analysis only examines spectral variability, only variable systematics or those which affect the analysis itself. This analysis assumes that there are no systematic differences in the source spectrum between the observations, as we use a mean spectrum calculated using all the available data. We test this assumption by analysing each observation independently, which returns the same result as the full analysis, although less detailed and degraded by the increased noise. A final concern is that PCA itself should only be applied to data which can be described as a linear sum of components, and will not return valid results from more complex systems. In these cases, independent component analysis (ICA) should be used instead, although this does not appear to be necessary for this analysis, as coherent results are returned.

\section{Results}

\begin{figure}
\centering
\includegraphics[width=\linewidth]{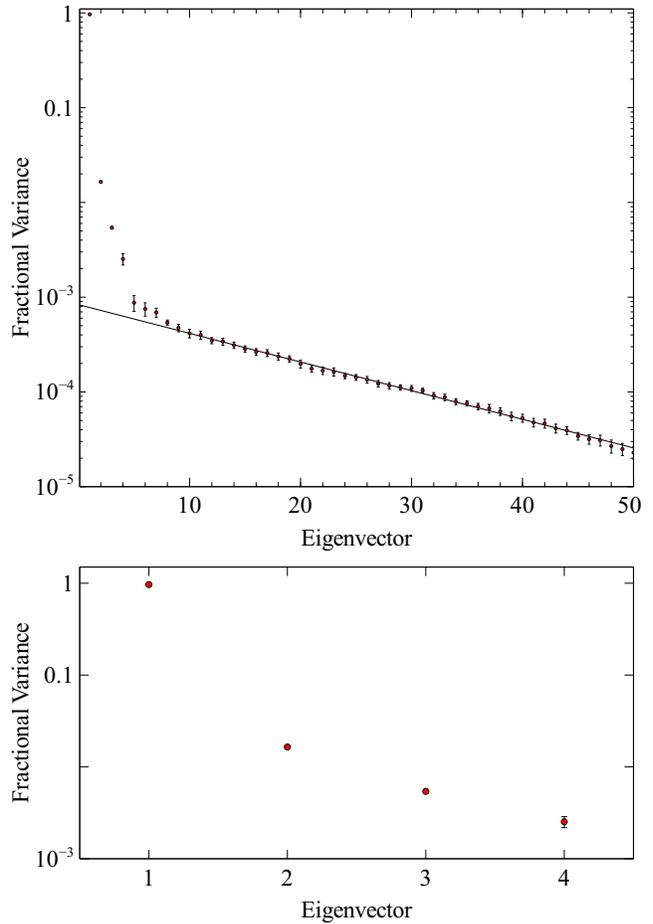}
\caption{Top: Log-Eigenvalue (LEV) diagram, showing the fractional variance of the first 50 eigenvectors from the PCA analysis of 600ks of XMM data on 10ks timescales. The line shows the best-fit geometric progression to the points from 5--50, which are attributable to noise. The significance of the components can be determined by their deviation from this line, which is $>5\sigma$ for all of the first four principal components. Note that the number of eigenvectors is either the number of energy bins, or the number of spectra, whichever is lower. Some higher order eigenvectors are excluded from this plot, for illustrative purposes.
Bottom: LEV diagram of the first four components. Error bars are plotted, but are smaller than the points for the first three components.}
\label{LEV_10ks_1}
\end{figure}

Fig.~\ref{LEV_10ks_1} shows the LEV diagram, also known as a scree diagram, for the PCA of the whole dataset, using 10~ks intervals. This is an effective way of quantifying the number of significant principal components, as well as the amount of noise, both in the analysis as a whole and within each component spectrum. The fractional variance found in components that are due to noise falls off geometrically \citep{Koljonen13,Jolliffe02}, and so lies along a straight line on the LEV diagram. We fit a geometric progression to the variance of the higher-order components (numbers 5 to 50), and use this to judge the significance of the first few components. We calculate the errors on the fractional variance using a similar method to that discussed in \citet{Miller07}, by randomly perturbing the input spectra, and finding the standard deviation of the variances of the resultant components.

\begin{figure}
\includegraphics[width=\linewidth]{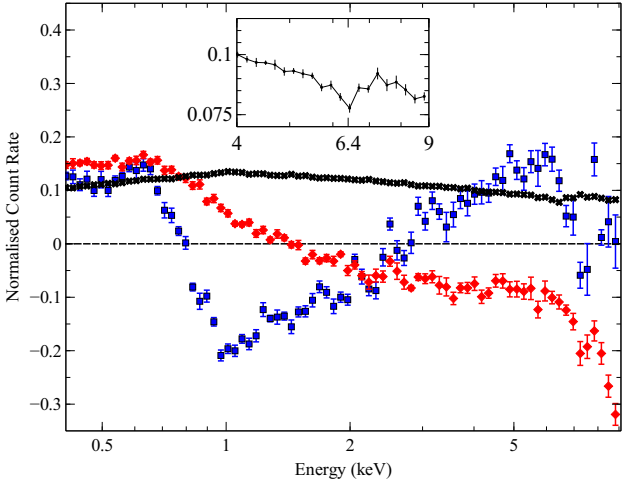}
\caption{Normalised spectra of the three significant components of the XMM analysis between 0.4 and 9keV (9-10keV is excluded, due to the increased noise that results in the weaker components). The first component is shown in black, the second in red, and the third in blue. Note that the PCA does not specify 
whether components should be positive or negative, since these represent deviations from the mean, they are just as likely to be either. Because the spectra have been normalised, they should not be interpreted as showing the exact spectral shape of each component, rather they show the strength of the correlations of each component as a function of energy. The inset shows the 6.4 keV dip in the first component.}
\label{PCA_specs_10ks}
\end{figure}

As found by \citeauthor{Vaughan04} and \citeauthor{Miller08}, between 2keV and 10keV using only the data from 2001, we find only one significant variable component, and all others are consistent with noise. However, including the full energy range or using all of the available data, we find that three components are needed to describe $\sim$99 per cent of the variability, on timescales between 1 and 20~ks. A fourth component is also found to be significant on 10ks and greater timescales when including all the data over the full energy range, however the variability in this component is small ($<0.5$ per cent), and cannot be distinguished from noise at shorter timescales. This means that it is impossible to impose a strong constraint on the variability timescales, and hence size of the emission region, for this component.Some or all of these components are presumably the same as those found by \citeauthor{Miller08}, although we cannot be certain, as the higher order components are not presented in that work. 

In the 10~ks analysis, all four of these components are found to deviate from the expected level of noise by $>5\sigma$, and are thus highly significant. More generally, we note that any components that are dominated by noise should not preferentially correlate between adjacent energy bins, or deviate significantly from the zero point in more than a few bins. Thus the probability of a component arising due to noise with significant deviations from the mean correlated between adjacent bins is vanishingly small. Finally, the distribution of normalised count rates within the spectrum of a principal component which is due to noise should be approximately Gaussian, with a mean of zero. We therefore check how well this distribution can be fit with a normal distribution, for each of the first 50 components. Using this method, we find good fits to the higher order components ($n>5$) with normal distribution, rule out such a distribution for eigenvectors 1--3 at high confidence levels, and find a relatively poor fit to eigenvector 4, although not sufficient to rule out this distribution.

The lower limit on the timescales involved is the size of the time intervals used to extract spectra. There is a trade-off between the number of bins (and hence temporal resolution) and the amount of noise, which can be quantified using the LEV diagram. As mentioned above, decreasing the interval size below 10~ks to 5 or 1~ks means that the fourth principal component is no longer significant.

The normalised spectra of the three most variable components are shown in fig.~\ref{PCA_specs_10ks}. Errors are calculated using the method discussed in \citet{Miller07}, where the input spectra are randomly perturbed, and the pricipal components recalculated. The errors are then found from the variance of the resultant components. We note that it is possible using this method that differences in the normalisations of the resultant components will contribute to the final estimate of the errors, but will not change the spectral shape, potentially leading to overestimated errors.

The first component is relatively flat, decreasing at low and high energies, and with a marked dip around 7keV. Because the PCA analyses variation from the mean, and the powerlaw makes up a smaller fraction of the observed flux at these energies, where the reflection spectrum is larger, this is what would be expected in the case of a varying powerlaw component, consistent with the results from both previous work with PCA and other methods \citep{Vaughan04,Miller08}. 

The second component is more complex. Values below zero imply an anticorrelation between those energy bins
and the positive bins, and there is a clear anticorrelation between low and high energies. This is very similar to the second component seen in the analysis of Cygnus X-1 by \citet{Malzac06}, who found that this component could be attributed to pivoting of the spectrum. 

\begin{figure}
\includegraphics[width=\linewidth]{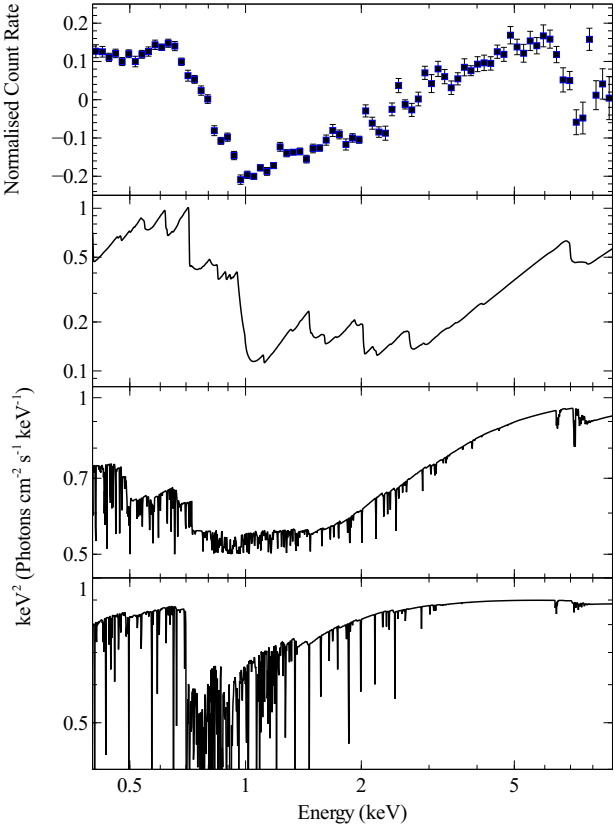}
\caption{Spectra of the third principal component (top), blurred reflection (upper middle), partial covering ionised absorption of a $\Gamma=2$ powerlaw, and absorbed powerlaw (bottom) from the best fitting broadband model (XMM-NuSTAR, Marinucci et al. in prep.) dominated by the warm absorber, shown from 0.4--9~keV, to coincide with the PCA results. Qualitatively, it appears that the shape of the third principal component is a closer match to the spectral shape of the reflected emission. The minimum flux in the principal component spectrum matches closely with the minimum in the blurred reflection spectrum, and the strong iron-line like feature is clearest in the reflection spectrum. We test these models quantitatively in section~\ref{extremes}.}
\label{reflection_absorption}
\end{figure}

The third component shows the characteristic soft excess and broad iron line of a reflection spectrum.
However, there is a dip at intermediate energies, implying an anticorrelation. This is likely to be due to the nature of PCA itself, as any variation will be fit initially by the first principal component. Therefore, if the normalisation of the reflection component were to change, it would be fit with the powerlaw first, then a correction would be applied to the powerlaw shape to make it look like reflection. At intermediate energies, where the spectrum is powerlaw dominated, this leads to an anticorrelation in the normalised spectrum. However, it is also possible that this is due to a real change in the shape of this component (see \S\ref{discuss}). The point between 7 and 8 keV is likely to be anomalous. The exact shape of the last few bins in this component seems to depend strongly on the amount of data used, and the number of bins. We interpret this as the effects of noise leaking into this component, and note that the bins appear to converge towards zero as more data is added to the analysis. Fig.~\ref{reflection_absorption} compares the spectrum of this component with those of the absorbed powerlaw and relativistic reflection from the model (Marinucci et al., in prep.). It is obvious from this that the third component is much closer to the blurred reflection spectrum, in terms of spectral shape, and cannot be explained by changes in the warm absorber. We note that the peak of iron line in the principal component spectrum is at a lower energy than that in the model spectrum, and we attribute this to the effect of dilution by narrow lines from the neutral reflection.

The fourth component also shows an excess around the iron line, but no soft excess. As the smallest and
noisiest component, it is the hardest to analyse. Using the expected noise level for the fourth component calculated using the LEV diagram, we estimate that the signal to noise ratio in this component is less than $\sim2$. It is conceivable that this component represents a change in the properties of the reflection spectrum, variations on long timescales from a distant reflector, or a change in the properties of the absorption. For the remainder of this work we restrict out analysis to the three main components, which can be more thoroughly investigated.

\subsection{Fitting extremal spectra}
\label{extremes}
\begin{figure}
\centering
\includegraphics[width=\linewidth]{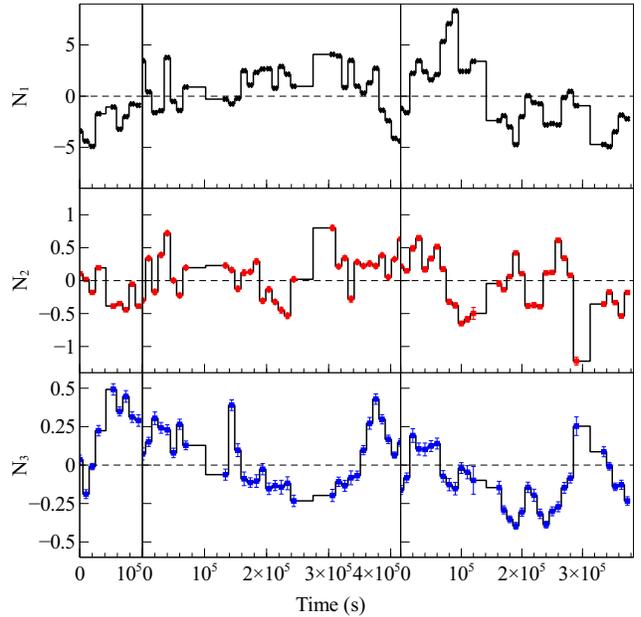}
\caption{Normalisations, $N_\textrm{i}$, of the first three significant components plotted against time. These are effectively lightcurves for each principal component. Error bars are plotted for all three, but are smaller than the points for component one and approximately the same size as the points for component 2. 10~ks intervals are used here, and all three components show significant variability on this timescale. Vertical lines separate the observations from different years.}
\label{pclcurves}
\end{figure}

Because of anticorrelated bins, it is impossible to fit models directly to the second and third components. The normalised spectra must be multiplied by the mean spectrum to convert to physical units, meaning that the anticorrelated bins give negative count rates which cannot be fit in \textsc{Xspec}.
However, we can investigate the effects of the component spectra on the mean source spectrum. We create a simple model, comprised of a linear combination of the three main principal components. We then fit this model to each of the normalised variation spectra used to calculate the PCA. This returns a continuous set of normalisations for each component, effectively the same as a lightcurve for each component. These normalisations can then be used to investigate the behaviour of each component, and the effect it has on the spectrum, in more detail. These lightcurves are shown for the first three components in Fig.~\ref{pclcurves}

We use the sets of normalisations to find the minimum and maximum normalisations of each component. These can then be used to find corresponding spectra. The normalised spectra output from the PCA can be converted to `real' spectra by multiplying by the mean spectrum, and normalisation value of that component:
\begin{equation}
F_\mathrm{i}(E_\mathrm{j})=N_\mathrm{i}\times F_\mathrm{mean}(E_\mathrm{j}) \times f_\mathrm{i}(E_\mathrm{j})
\end{equation}
where $E_\mathrm{j}$ are the energy bins, $f_\mathrm{i}(E_\mathrm{j})$ is the normalised component spectrum and $N_\mathrm{i}$ are the normalisations for each component. 
These spectra show the deviations from the mean, in counts s$^{-1}$, caused by
variations in the $i$th component. These spectra will still contain negative values for the second and third 
components, so still cannot be modelled trivially. However, by adding the minimum and maximum spectra for each component to the mean spectrum of the whole dataset
we can generate extremal spectra for each component:
\begin{equation}
F_{\pm\mathrm{,i}}(E_\mathrm{j})=N_{\pm \mathrm{,i}}\times F_\mathrm{mean}(E_\mathrm{j})\times f_\mathrm{i}(E_\mathrm{j})+F_\mathrm{mean}(E_\mathrm{j})
\end{equation}
where $F_{\pm \mathrm{,i}}$ and $N_{\pm \mathrm{,i}}$ are the maximum and minimum spectra and normalisations, respectively, 
for each component. These spectra correspond to the source spectrum when the $i$th component
is at an extreme value. Because these extremal spectra are all non-zero, they can safely be imported into 
\textsc{Xspec} for modelling.
\begin{figure}
\centering
\includegraphics[width=\linewidth]{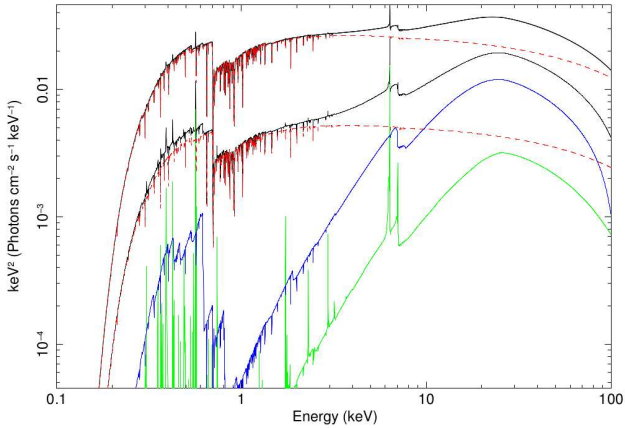}
\caption{Model used to fit the minimum and maximum spectra generated with PCA component 1. Blurred reflection (blue)
and distant reflection (green) are kept constant, and the powerlaw (red,dashed) is allowed to vary between the 
two spectra. Full models are shown by black solid lines.}
\label{powerlaw_model}
\end{figure}
\begin{figure}
\centering
\includegraphics[width=\linewidth]{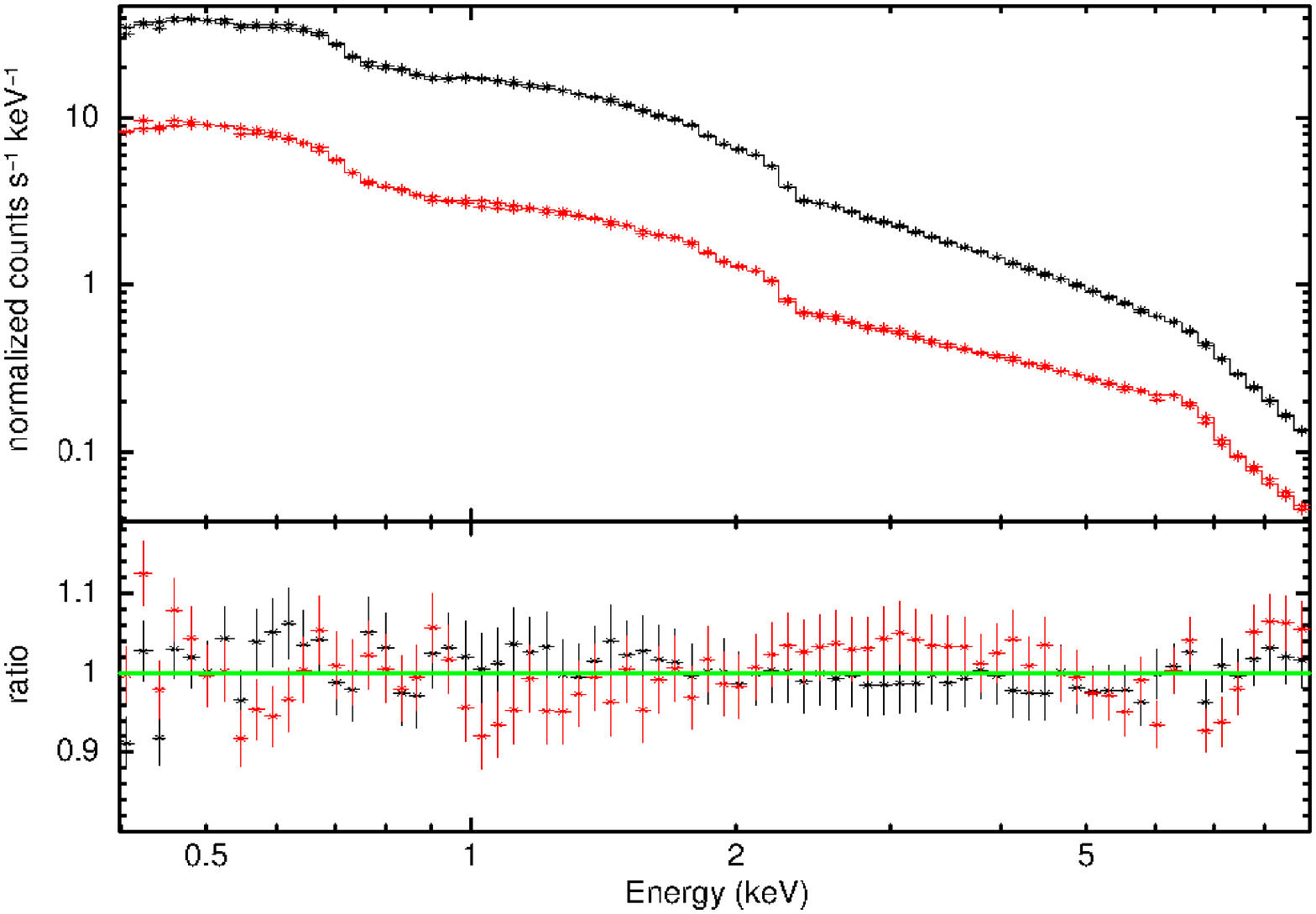}
\caption{Folded data and residuals from fitting the broadband model (XMM-NuSTAR, Marinucci et al. in prep.) to the minimum (red) and maximum (black)
spectra for the first principal component. Only the powerlaw normalisation is allowed to vary between the the two spectra.
}
\label{comp1_minmax}
\end{figure}

To test our interpretation of the three main components, we fit the minimum and maximum spectra simultaneously,
with the model described in Marinucci et al. (in prep.). This model includes a powerlaw, both relativistic and distant reflection 
\citep[both modelled with \textsc{Xillver}, and convolved with \textsc{Relconv} for the relativistic blurring, see][]{Garcia13,Dauser10}, and two absorbing zones.
For component 1, we allow the powerlaw normalisation to vary between 
the two spectra, keeping all other parameters the same. Using this method, we
obtain a $\chi^2$ value of 130/148, with powerlaw normalisations of $(3.5\pm0.1)\times 10^{-2}$ and
$(6.40\pm0.25) \times 10^{-3}$.
The model for this is shown in fig.~\ref{powerlaw_model}, and the data and residuals in fig.~\ref{comp1_minmax}. We note that
there are some residuals not accounted for perfectly by varying the powerlaw alone, particularly around 6--7~keV, and suggest
that this might be due to some extra component being weakly correlated with the powerlaw, and ``leaking" in to the first PCA 
component. This could be caused by intrinsic variations in the flux from the corona, which would be correlated with the 
reflected emission.

\begin{figure}
\centering
\includegraphics[width=\linewidth]{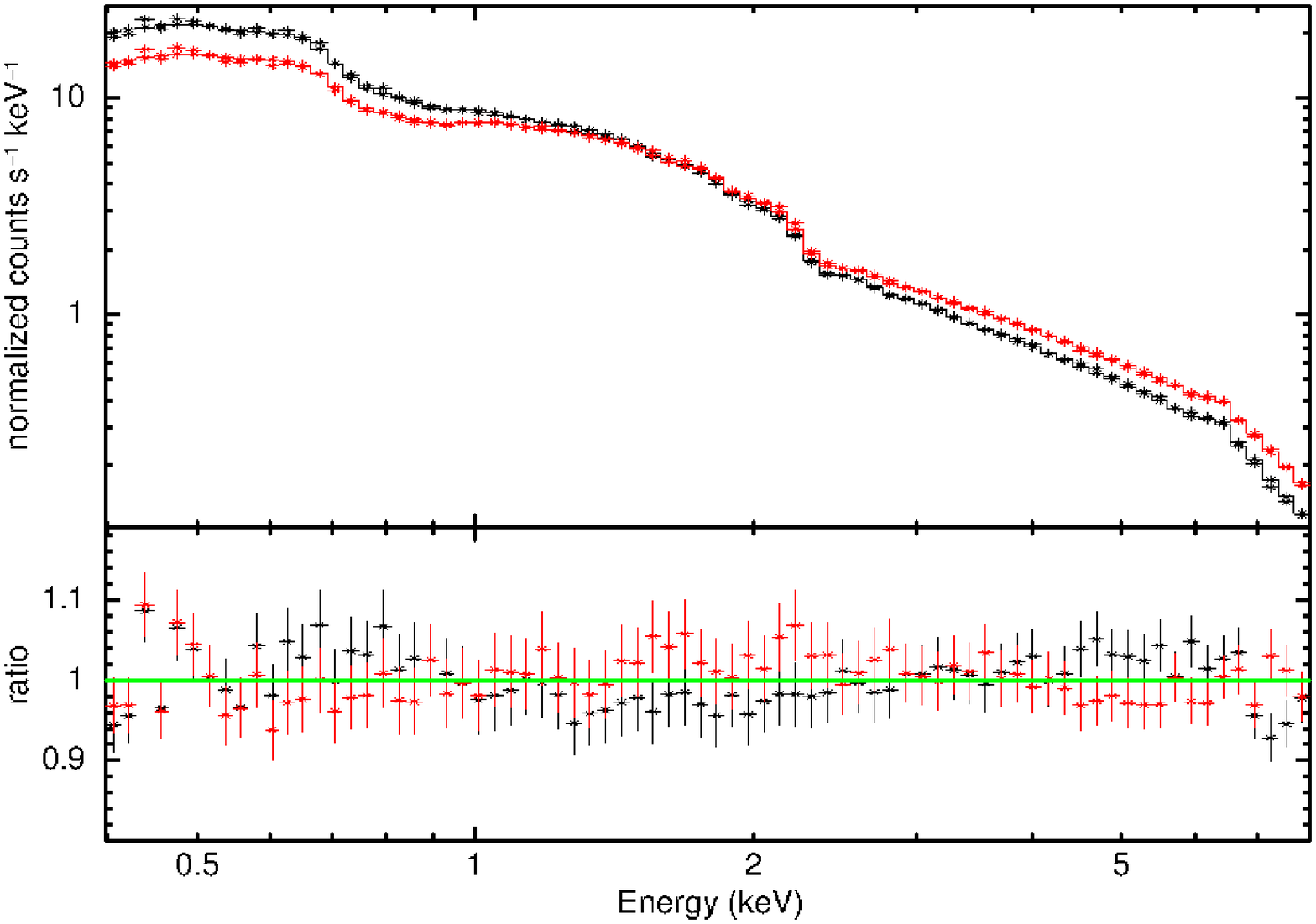}
\caption{Folded data and residuals from fitting the broadband model (XMM-NuSTAR, Marinucci et al. in prep.) to the minimum (red) and maximum (black)
spectra for the second principal component, allowing the photon index and powerlaw normalisation to vary between the the two spectra.}
\label{comp2_minmax}
\end{figure}
For component 2, allowing the powerlaw normalisation and index to vary between the minimum and maximum 
spectra (shown in fig.~\ref{comp2_minmax}) gives a $\chi^2$ value of 126/129. The normalisation changes
between $(1.39\pm0.03)\times 10^{-2}$ and $(1.64\pm0.03 \times 10^{-2}$, and the index varies between $1.82 \pm 0.01$ and 
$2.08 \pm 0.01$. Note that the increase in powerlaw normalisation should not be confused with an increase in flux,
as the powerlaw gets steeper at the same time, causing a net reduction in the count rate. It is also possible that 
a change in the absorption could be responsible for some or all of the variability in this component. We test this
by adding a partial covering absorber to the model, using \textsc{zxipcf} \citep{Reeves08}, and allowing the covering fraction and column density to vary between the two spectra. 
We find that the differences below 2keV can be explained adequately using this model, but that above 2keV the fit is poor, giving a $\chi^2$ of 547/128. By freeing up the powerlaw normalisation, the fit can be improved to give a $\chi^2$ of 145/127, which is a large improvement, although still significantly worse than the pivoting model.

\begin{figure}
\centering
\includegraphics[width=\linewidth]{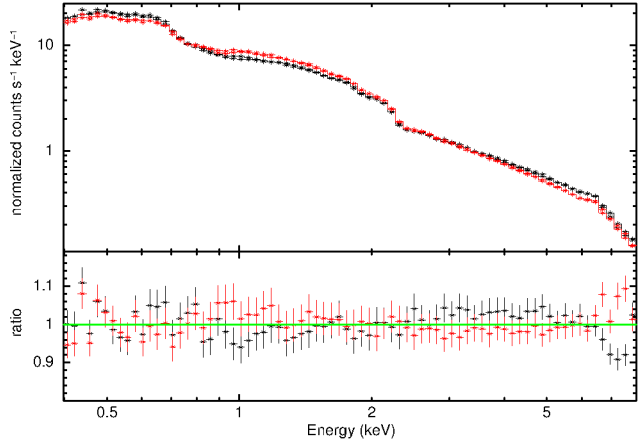}
\caption{Folded data and residuals for the minimum and maximum spectra of the third principal component, fit by allowing both powerlaw
and reflection normalisations to change.}
\label{comp3_minmax}
\end{figure}

We initially fit the third component by allowing only the blurred reflection normalisation to vary between the minimum and maximum spectra. This gives an unacceptable reduced $\chi^2$ of 190.2/128, with the normalisation varying between $4.0\times 10^{-5}$ and $7.4\times 10^{-5}$. This is unsurprising, given the anticorrelated bins at
intermediate energies, which cannot be fit with a single additive component. We test two different explanations for these bins: firstly, if the variability in the blurred reflection component is being fit initially by the first PCA component, as seems likely because of the nature of the analysis, the third component should appear as a 
correction factor to the first, rather than as the exact spectrum of the reflection component. Thus, an excess appears where the reflection component is larger, and a deficit where is it smaller. To test this, we allow the powerlaw 
normalisation to vary as well, which gives a much lower $\chi^2$ of 127/127 with the powerlaw normalisation changing by a factor of $\sim10$ per cent, and the reflection normalisation increasing from $4.7\times 10^{-5}$ to $1.1\times 10^{-4}$. The
fit to this model is shown in Fig.~\ref{comp3_minmax}. Secondly, if the change in the spectral shape is intrinsic to the source, it is possible that the anticorrelation could be caused by a change in the ionisation parameter, $\xi$. Allowing $\xi$ to vary between the two spectra gives a $\chi^2$ of 138/127, and allowing \emph{both}
$\xi$ and the powerlaw normalisation to very gives 118/126. 
We also attempt to fit these spectra by adding a partial covering absorber to the model, using \textsc{zxipcf} . Allowing only the covering fraction and column density to vary between the two spectra gives a much worse fit than varying the reflection normalisation alone ($\chi^2_\nu=480/128$), nor can we achieve a good fit ($\chi^2_\nu=440/129$) when we allow the powerlaw normalisation to change as well (which does not necessarily imply correlation, as discussed above). Although a partial covering absorber is qualitatively similar in terms of spectral shape, it cannot quantitatively explain the variability in this component, as the flux is not sufficiently altered at low energies to explain this variability.

It is likely that, given the large number of time intervals used and the relatively large errors on the normalisations of the weaker spectral components, the extreme normalisations of the principal components are exaggerated. The best fit values should not therefore be taken as typical for the source behaviour, which is likely to be more conservative.

\subsection{Modelling}

We now use spectral fitting as a complementary and independent (but not model independent) way of tracking
the variable components of the spectrum of MCG--06-30-15. We fit the broadband model of Marinucci et al. to each of the 
input spectra.
\begin{figure*}
\centering
\includegraphics[width=\linewidth]{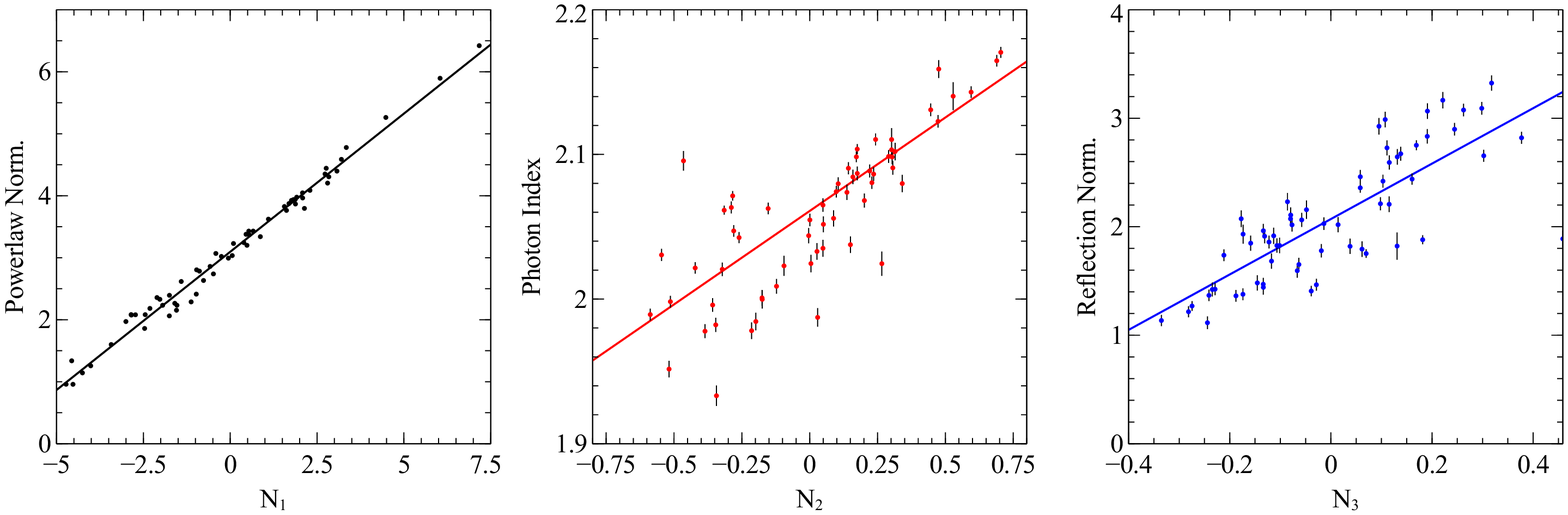}

\caption{Correlations of the normalisations, $N_\mathrm{i}$, of the components extracted by the PCA with the parameter values from fitting a reflection plus powerlaw model to the same set of spectra. Component one (black) is plotted against the powerlaw normalisation, component two (red) against the photon index, and component three (blue) against the blurred reflection normalisation.
The same set of spectra are used for both the PCA and spectral fitting, with 10ks time bins, and all normalisations are plotted linearly. The model is that described in Marinucci et al., in prep., shown in fig~\ref{powerlaw_model}, with all parameters frozen except for the normalisations and photon index. The powerlaw and reflection normalisations are multiplied by factors of 100 and 10000, respectively, for clarity.}
\label{PCA_correlations}
\end{figure*}

We use \texttt{PyXspec} 1.0.1 to fit the model to the full set of input spectra used in the PCA, from all XMM observations, leaving only the powerlaw and reflection normalisations and photon index free. We then compare the values of each parameter from these fits with the normalisations of the principal components, and find a strong correlations between component one and the powerlaw normalisation, component two and the powerlaw index, and component 3 and the reflection normalisation.
The correlation coefficients are 0.996, 0.75 and 0.83 respectively, and the probabilities of these correlations arising by chance are all less than $10^{-12}$. These correlations are shown in fig.~\ref{PCA_correlations}.

From this we conclude that the PCA components correspond to the model powerlaw normalisation, index
and reflection normalisation, and that the PCA can then be used as a tool to investigate the properties of these parameters, although we note that it is entirely possible for the PCA components to be related to more than one physical component, for example if the reflection and powerlaw normalisations were correlated, this would show up as a single principal component.
We investigate the timescales of the variations in each component by restricting the PCA to look only at
variations between adjacent bins, and then perform the same analysis with 1ks bins. All three components are found to have significant variablility down to 1ks timescales using PCA, and we confirm this using spectral fitting.
Timescales smaller than this are harder to probe, due to the increase in noise. 

We find no correlation between any pair of components from the PCA, or between the model parameters from fitting
the spectra. While the PCA does, by definition, look at independently varying components, this does not require the 
components to be uncorrelated \citep[see e.g.][]{Koljonen13}, so long as they have enough independent variation to 
distinguish them as separate. 

\section{Discussion}
\label{discuss}
It is highly significant that all three PCA components are found to vary on timescales less than 1~ks as this can be used to constrain the size of the responsible emitting region.
The mass of MCG--06-30-15 is not very well constrained, largely because it is too far away for the sphere of influence 
around the black hole to be resolved. A recent study by \citet{Raimundo13} found an upper limit on the mass of 
$6\times 10^7 M_\odot$ using stellar dynamics; \citet{Vasudevan09} use the method of \citet{Mushotzky08}, based 
on the relation between the K-band luminosity and black hole mass to find a value of $M_\mathrm{BH}=1.8\times 10^7M_\odot$ and
\citet{McHardy05} and \citet{Bennert06} use multiple methods to find masses in the range 3 to 6$\times 10^6M_\odot$ and 
$8\times 10^6$ to $2.7\times 10^7M_\odot$, respectively.

The light travel time over
one gravitational radius, $R_\mathrm{G} = GM/c^2$, is given by $R_\mathrm{G}/c$, so the size of the emitting region in $R_\mathrm{G}$ must be less
than
\begin{equation}
R=t_\mathrm{var}\times \frac{c}{R_\mathrm{G}}=t_\mathrm{var} \times \frac{c^3}{GM}
\end{equation}
where $t_\mathrm{var}$ is the variability timescale.
If we consider masses in the range from $3 \times 10^6$ to $3 \times 10^7 M_\odot$ and $t_\mathrm{var}=1000$~s, this constrains
the emitting region to be $R \lesssim 7 \text{--}70~ R_\mathrm{G}$. This is consistent with our interpretation of the 
first two components as continuum emission from a compact corona, close to the event horizon, and the third as variations in a
relativistically blurred reflection component. The size and location of the corona in rapidly accreting black holes 
has previously been constrained by studies of time lags \citep[see e.g.][]{Wilkins13,Kara13,Zoghbi10}, microlensing \citep[e.g.][]{Dai10}
and emissivity \citep{Walton13}, which consistently find that the X-ray emission region must be small and close to the event horizon \citep[See][for a summary]{Reis13}.

We note that although MCG--06-30-15 does show evidence of discrete absorption events, we do not find a component that is well fit by absorption alone. In the latest observation, there are several time intervals with an unusually high hardness ratio and low flux, which are interpreted as a cloud passing over the line of sight (Marinucci et al., in prep.). It is likely that unusual events such as this are fit by the PCA using some combination of the principal components, and thus a separate component is not found. By investigating the component normalisations during these intervals, we find that the majority of the variation in these bins is fit with an extremely low powerlaw normalisation (component one), rather than any major changes in the higher order components.

In previous work \citep[eg.][]{Fabian02,Vaughan03}, the reflection component was found to be relatively constant, when compared to the powerlaw component however,in the spectral fitting analysis, shown in fig.~\ref{PCA_correlations}, we find that both the blurred reflection and powerlaw normalisations vary by a factor of $\sim4$.  We do not believe this to be in conflict with previous results, for several reasons:
firstly, because the magnitude of the powerlaw component (Fig.~\ref{powerlaw_model}) is much larger than
that of the blurred reflection component it dominates the variability of the spectrum; secondly, the variations in the blurred reflection spectrum will be diluted by the presence of distant, neutral reflection, lowering the variable fraction in the reflection dominated bands; and finally because there is a large scatter in the results from spectral fitting, caused partly by noise, and partly by fitting data from over a decade, including absorption events, with only three parameters. We confirm this by looking more closely at spectra found to have the most extreme reflection normalisations, and find that, when modelled more carefully, more conservative parameters are favoured. 

If the blurred reflection component is dominated by emission close to the inner edge of the disk, as seems likely from the variability timescales, it is plausible that a non-uniform disk surface could cause relatively rapid variations in the ionisation parameter, which is one possible explanation of the anticorrelated bins in the third PCA component. The degeneracy between ionisation and powerlaw normalisation changes could be broken by examining the PCA component spectra from other sources with different reflection spectra, and exhibiting different variable behaviour, or by using broad band time resolved spectroscopy to examine the spectral shape of the components more precisely.

There is scope for using PCA on a set of lightcurves rather than spectra in future work. By using 
lightcurves from relatively broad energy bins, it should be possible to investigate very rapid variability using PCA, rather than being restricted to $\gtrsim$1~ks timescales. We note that with higher count rate spectra it is possible to probe smaller timescales, such as the analysis of Cygnus X-3 by \citet{Koljonen13} who investigate timescales as low as 1 minute. Within this work we consider intervals on the scale of hours and days, and on the scale of years (although there appear to be no significant differences in the spectral of MCG--06-30-15 between the observations in different years). Any potential variations on month-long timescales cannot be investigated with the available \emph{XMM-Newton} data, and may be hiding variations in other spectral components, particularly the warm absorption.

It is interesting to consider that PCA, once the input spectra have been properly normalised, should be largely independent of the instrument used to find the spectra. This raises the possibility of combining data from different instruments, with careful binning and selection of time intervals, into a single analysis. However, this is quite definitely beyond the scope of this paper.

There is a possibility that the components from the PCA contain `hidden', weak correlations between the spectral components, which could be revealed by more extensive modelling. For example, if the reflection component is weakly correlated with the powerlaw, it could be that the first component would be well described by a powerlaw plus a small amount of relection, with a separate reflection component as well to describe the independent variability. This could be distinguished by fitting models to the extremal spectra, although we find adequate fits by only having one parameter free.

\section{Conclusions}

Using principal component analysis, we find that over 99 per cent of the variability of MCG--06-30-15 can be described by just 
three components. We calculate extremal spectra for these components, and find that they are well fit by: i) a change in 
the normalisation of the powerlaw; ii) a change in the photon index of the powerlaw; and iii) variations in the normalisation
of a blurred reflection component.

We confirm these results by comparing the normalisations of the PCA components with the parameters obtained from spectral
fitting, and find very strong correlations between the relevant fit parameters and the magnitudes of the components.
By comparing the PCA results using different time bins, we find that all three major components are variable down to 
timescales as low as 1000~s, which corresponds to a size of less than 7--70~$R_G$, depending on the mass of the 
black hole. This is consistent with our interpretation of the components as being due to changes in the position and intrinsic flux 
of a hot corona, close to the event horizon, and the reflected emission due to coronal photons hitting the accretion disk.

\section*{Acknowledgements}
MLP would like to thank Dan Wilkins and Roderick Johnstone for help with numerous technical 
issues and J. Malzac for helpful discussion, and acknowledges financial support from the 
Science and Technology Facilities Council (STFC). 
AM acknowledges Fondazione Angelo Della Riccia for financial support. 
ACF thanks the Royal Society for support.

\bibliographystyle{mnras}
\bibliography{mcg6_paper}
\end{document}